# Reduction of Excess Capacity with Response of Capital Intensity

Samidh Pal[1]

**Abstract:**

**Purpose:** The objective of this research was to show the response of the potential reduction of excess capacity in terms of capital intensity to the growth rate of labor productivity in the manufacturing industrial sector.

**Design/Methodology/Approach:** The research was carried out in 2019 in 55 groups of Indian manufacturing industry within six major Indian industrial states. Mainly, the research used the modified VES (Variable Elasticity Substitution) estimation model. The research focused on the value of the additional substitution parameter of capital intensity ($\mu > 0$).

**Findings:** Almost all selected industry groups with in six states need capital-intensive production. The results found additional parameter of capital intensity ($\mu$) is greater than zero for all industry groups. It means that a higher product per man can be obtained by increasing the capital per worker.

**Practical Implications:** Research shows that an increasingly need for capital investment in need for higher labor productivity is likely to induce the manufacturing unit to use more capacity in existence. It reveals that investors in these selected six states can increase their capital investment.

**Originality/Value:** The analysis of the result allowed to determine the fact that capital intensity is an essential variable for reduction of excess capacity which cannot be ignored in explaining productivity.

---

[1] Pursuing Ph.D, Department of Economic Science, University of Warsaw, s.pal@uw.edu.pl;

## 1. Introduction

A higher amount of capital per man may raise the productivity of man not only by quantity but also by quality of production. The elasticity of substitution between capital and labor shows how the ratio of production factors alters when their marginal rate of technical substitution changes by one percent (Thach et al., 2020). Therefore, the estimation of another production function requires the observation of capital investment for capacity utilization.

Hicks noted that in the absence of technological change, and increasing capital-labor ratio, might induce a tendency toward a diminishing of substitution, thereby at least implicitly recognizing the possibility of a variable elasticity of substitution (Lovell, 1968). It is found that when shareholders save more than workers or the substitution elasticity is higher than one, a country characterized by production functions with higher elasticity of substitution experiences higher equilibrium levels of capital and output per capita (Grassetti et al., 2018). This statement seems to have encouraged us to introduce the Variable Elasticity of Substitution (VES) production function of one form or another, designating it a homothetic or transcendental production function. However, several alternative methods are fitting the VES function. In this study, the method is derived from the original CES function, based on which the VES function is derived from. Therefore, our VES function is a more generalized function.

In our assignment, we are interested in reducing the excess capacity within six major industrial states. Operating an industrial enterprise considerably below capacity (one or two shifts when three would be more economic or without a great concern for cost control or efficiency or without any concern regarding the average productivity of labor) is evidence of excess capacity. There is a vast difference between this kind of operation and full capacity production at high efficiency. Usually, it amounts to the difference between substantial losses and high profits. Through the VES function, we deserve that when the additional substitution parameter of capital intensity for an industry group is greater than zero ($\mu > 0$), which means a higher product per man is obtained by increasing the capital per worker, or in other words by adding a more capital-intensive process of production in that industry group.

West Bengal is one of the least industrialized states which is in crisis process so far as the underutilization concerned where most industries are labor intensive and have greater wastage of capital. Furthermore, the centralized industries in that state face the problem of excess capacity. We considered the study of the potential reduction of excess capacity through a comparative study. From this comparative study, we will identify that which industry groups are needing capital intensive production urgently and how much more capital intensity they need to increase for their capital-intensive production. Moreover, we will compare the present conditions of West Bengal with other five states through an exploration with an industry group.

## 2. Literature review

Several Indian economists did some comparative empirical studies on capacity utilization for several Indian manufacturing industries. We collected some literatures about capacity utilization and discussed their used methodology and findings. Through these literature studies, we will try to identify the utilization of capacity with the optimum level, which will help us to reduce the problem of excess capacity.

Srinivasan et al. identified the presence of excess capacity for different Indian industries in terms of demand and supply constraints. Through his empirical research study, the results found that most of the considered Indian industries are demand constrained. In addition, he described in his research outcomes that the estimated shortage of production capacities has not been uniform or same over the years (Srinivasan, 1992).

Gajanan et al. estimated the capacity utilization rate for some selected Indian manufacturing industries through the generalized Leontief variable cost function with fixed capital input. They found that the capacity utilization was higher before the 80s, then dropped in after 80s, and then again started rising in the early 90s. In their study, they also found that the capacity utilization is positively related to the labor intensity of production, which holds both between-industries and within-industries. Their empirical results also identified that the traditional measures of capacity utilization for minimum output-capital ratio are not appropriate for short run decision making of a firm (Gajanan & Malhotra, 2007).

Kumar et al., they analyzed the regional variation of excess capacity of Indian sugar industry. They used the time series data from 1974 to 2005 and they provided the trends of capacity utilization from a regional perspective. They used a methodology that is not so famous. They used a linear programming-based method DEA (data envelopment analysis). As per their findings, they said that the capacity utilization fell mainly because of untimely payment for the purchase of raw materials, lack of labor inputs, and excessive government controls. In addition, they explained that if the increased capital per man remains underutilized owing to some other cause (such as lack of raw materials) then the excess capacity will be increased in the representative manufacturing firm (Kumar & Arora, 2009).

A. K. Deb tried to identify the channels through which economic reforms enhanced the productivity growth in total Indian manufacturing industry. In his empirical study, he used the subsequent regression analysis. For the utilization of a plant's capacity, he estimated the capacity utilization rate. Over the post-reform period, the annual average capacity utilization rate is lower in Indian manufacturing industries. Through his research, the main finding is that the capacity utilization rate increased faster after the reforms (after 2001 - 2002) in most of the major industrial states. He also confirmed through the subsequent regression analysis that there was evidence of a favorable impact of economic reform on productivity growth in total manufacturing sector, beyond the positive impact of improved capacity utilization (Deb, 2014).

The above listed literatures are still not satisfactory in some points of view. Because either they expected the causes of excess capacity, or they just found the tendency of capacity utilization. However, in our research study, we made a comparative study to identify which industry groups needs how much urgent capital investment to increase the capacity utilization through capital intensive production. Exception of our study is that we took all 55 industry groups together and then we performed a GLS (generalized least square) model which we get after linearizing the VES function. Then we manually calculated the parameter μ from the coefficient of GLS model, which is an additional substitution parameter of capital intensity. With this additional substitution parameter, we made a comparative analysis with the elasticity of substitution parameters of CES functions of all 55 industry groups. In this comparative study, we also compared the industry group-wise present situation in West Bengal, which is a specialized state.

### 3. Measure of capital intensity

The VES production function explicitly permits the capital-labor ratio to be an explanatory variable of productivity. A comprehensive study by Hildebrand and Liu has shown the weakness of the CES production function in which the efficiency parameter has a downward bias because of omission of the variable (K/L), for which it does not explain the productivity variation in many industries. The VES production function overcomes this defect of the CES. Hildebrand and Liu* have suggested that "If one relies upon the goodness of fit of an empirical relationship as the initial basis for deriving a theoretical one, as Arrow, Chenery, Minhas, and Solow did, then probably would have to consider the three-variable relationship (*V/L, W and K/L; where W is wage*) as better established than the two-variables (*V/L and W*)."

$ln\ (V/L) = ln\ a + b\ ln\ W + c\ ln\ (K/L) + e$              (1.1)

Where c is a constant and the other notations are the same as before. When the production function $V = F\ (K, L)$ is homogeneous of degree one, we may rewrite in

$V/L = F\ (K/L,\ 1)$              (1.2)

        By setting in (1.2)   $V/L = Y$       and         $K/L = X$

Then, we have

$$Y = f(X) \tag{1.3}$$

or, $V = L f(X)$ (1.4)

Let $W$ be the wage rate with output as the numeraire in equation (1.1). If both labor and product markets are competitive, then

$$W = f(X) - X\acute{f}(X) \tag{1.5}$$

or, $W = Y - X \dfrac{dY}{dX}$ (1.6)

and $r = \acute{f}(X)$ (1.7)

where, $\acute{f}(X)$ is the marginal product of capital, $f(X) - X\acute{f}(X)$ the marginal product of labor, and $r$ is the returns to capital.

By substituting (1.1), we get the following differential equation:

$$\ln Y = \ln a + b \ln\left(Y - X\frac{dY}{dX}\right) + c \ln X. \tag{1.8}$$

Solving for $\dfrac{dY}{dX}$ results in

$$\frac{dY}{dX} = \frac{Y}{X} - a^{-\frac{1}{b}} X^{-\frac{c}{b}-1} Y^{\frac{1}{b}} \quad \text{(Bernoulli's equation)} \tag{1.7}$$

Divided by $Y^{\frac{1}{b}}$ both side of equation (1.7)

$$Y^{-\frac{1}{b}} \acute{Y} = \left(\frac{Y^{1-\frac{1}{b}}}{X}\right) - \left(a^{-\frac{1}{b}} X^{-\frac{c}{b}-1}\right) \tag{1.8}$$

$$\left(1 - \frac{1}{b}\right) Y^{-\frac{1}{b}} \acute{Y} = \left(1 - \frac{1}{b}\right)\left[\left(\frac{Y^{1-\frac{1}{b}}}{X}\right) - \left(a^{-\frac{1}{b}} X^{-\frac{c}{b}-1}\right)\right] \tag{1.9}$$

$$\left(Y^{-\frac{1}{b}}\right)' = \left(1 - \frac{1}{b}\right)\left[\left(\frac{Y^{1-\frac{1}{b}}}{X}\right) - \left(a^{-\frac{1}{b}} X^{-\frac{c}{b}-1}\right)\right] \tag{1.10}$$

Since the differential equation is nonlinear, it is difficult to solve. However, by letting

If, $z = Y^{1-\frac{1}{b}}$ putting in equation (1.10)

$$z' = \left(1 - \frac{1}{b}\right)\left[\left(\frac{z}{X}\right) - \left(a^{-\frac{1}{b}} X^{-\frac{c}{b}-1}\right)\right] \tag{1.11}$$

$$z' = \frac{b-1}{b} z \frac{1}{X} - \frac{b-1}{b} a^{-\frac{1}{b}} X^{-\frac{c}{b}-1} \tag{1.12}$$

$$z' + \frac{1-b}{b} z \frac{1}{X} = \frac{1-b}{b} a^{-\frac{1}{b}} X^{-\frac{c}{b}-1} \tag{1.13}$$

By linear nonhomogeneous first-order differentiation of equation (1.13)

$$z X^{\frac{1}{b}-1} = \frac{1-b}{1-b-c} a^{-\frac{1}{b}} X^{\frac{1-b-c}{b}} + \beta \tag{1.14}$$

$$z = \left[\frac{1-b}{1-b-c} a^{-\frac{1}{b}}\right] X^{\frac{-c}{b}} + \beta X^{1-\frac{1}{b}} \tag{1.15}$$

or, $z = \alpha X^{-\frac{c}{b}} + \beta X^{1-\frac{1}{b}}$ (1.16)

where, $\alpha = \left[a^{-\frac{1}{b}}\left(\frac{1-b}{1-b-c}\right)\right]$ and $\beta$ is the arbitrary constant of integration. By transforming z back to Y in (1.16) $z = Y^{1-\frac{1}{b}}$, we obtain a new production function.

$$Y = \left[\alpha X^{-\frac{c}{b}} + \beta X^{\frac{b-1}{b}}\right]^{\frac{b}{b-1}} \tag{1.17}$$

or, $V = \left[\beta K^{\frac{b-1}{b}} + \alpha \left(\frac{K}{L}\right)^{-\frac{c}{b}} L^{\frac{b-1}{b}}\right]^{\frac{b}{b-1}}$ (1.18)

Express equation (1.18) in the SMAC (Solow, Minhas, Arrow and Chenery) notation,

$$V = \left[\beta K^{-\rho} + \alpha \left(\frac{K}{L}\right)^{-\mu(1+\rho)} L^{-\rho}\right]^{-\frac{1}{\rho}} \qquad \text{Where, } \rho = \frac{1}{b} - 1 \qquad (1.19)$$

By setting $\qquad \alpha = (1-\delta)A^{-\rho} \qquad$ and $\qquad \beta = \delta A^{-\rho}$

We obtain

$$V = A\left[\delta K^{-\rho} + (1-\delta) \left(\frac{K}{L}\right)^{-\mu(1+\rho)} L^{-\rho}\right]^{-\frac{1}{\rho}} \qquad (1.20)$$

This new production function has the same form as the CES function except $L^{-\rho}$ which is multiplied by $\left(\frac{K}{L}\right)^{-\mu(1+\rho)}$. Obviously, if $\mu$ is equal to zero, the new function reduces to the CES function; therefore, the new function is a more general from which includes the CES function as a special case. To check the response of the additional parameter of substitution $\mu$, we linearized this VES function by first-order Taylor's expansion to the study of "Response" of capital intensity for the potential reduction of excess capacity. [*See in detail **Appendix: A**]

This function is stated in stochastic form. Here A, $\delta$ and $\rho$ are the efficiency, distribution, and substitution parameters, respectively. In this function, $\mu$ is an important parameter which holds an important explanation as when $\mu>0$, a higher product per man is obtained by increasing the capital per worker, or in other words, by introducing a more capital-intensive process of production needed (Kazi, 1978). When $\mu=0$, the VES function is transformed automatically into CES function.

The VES differs from the CES in one important aspect. The CES requires that the elasticity of substitution be the same at all points of an iso-quant, independent of the level of output, hence at all points of the iso-quant map. The VES on the other hand, requires that this substitution parameter should be the same only when the lines are from the origin (Revankar, 1971a). In this analysis, we are interested to focus on the response of capital. For this, we break the function manually to determine the $\mu$ and other parameters.

### 4. Selection of best generalized model

We took two industry groups (264 and 221) which are classified as centralized and decentralized industry groups, or they are classified on the basis of unequal distribution of capital intensity. For our linearized variable elasticity of substitution (VES) model, we selected three types of OLS models: 1) polynomial, 2) power, and 3) exponential. The polynomial and exponential model are coded in R with 'glm' function which is known as the generalized least square function and the power model coded with 'lm' function which is known as least square function. As per our mathematical interpretation, we already described that our VES model is a more generalized model which includes the CES function as a special case. After we linearized the VES function, the model took the polynomial shape (1.21). In all these three types of OLS models, the capital intensity is depicted as an explanatory variable.

Polynomial model: $\qquad \left[\ln \frac{V}{L}\right]_i = \beta_0 + \varphi_1 \left[\ln \frac{K}{L}\right]_i^1 + \varphi_2 \left[\ln \frac{K}{L}\right]_i^2 + \varphi_3 \left[\ln \frac{K}{L}\right]_i^3 + \cdots + e \qquad (1.21)$

Exponential model: $\quad \ln\left[\frac{V}{L}\right]_1 = \beta_0 + \beta_1 \left[\frac{K}{L}\right] + e \qquad (1.22)$

Power model: $\qquad \ln\left[\frac{V}{L}\right]_1 = \beta_0 + \beta_1 \ln\left[\frac{K}{L}\right] + e \qquad (1.23)$

To get the best generalized OLS model for the selected industry group wise, we calculated the standardized root mean square error (SRMSE) and compared those values. The interpretation of SRMSE is: if SRMSE < 0.5, then it means that good fit, if SRMSE ~ 0.75, then it means that decent fit, and if SRMSE > 1 then it means that bad fit and the existence of off-scale observations.

The result shows (Table: 1) that the polynomial model is good, because all subsequent powers of the explanatory variable are significant. Moreover, the value of the standardized root mean square error (SRMSE) of the polynomial model is less than 0.5, it means that the observations are fits well in the model. Based on the AIC (Akaike information criterion), it can be concluded that polynomial estimation (lower AIC) is significantly better than the exponential and power estimation (higher AIC). Therefore, we decided to perform our estimations for all industry groups with a polynomial model.

***Table: 1.*** *Industry code wise least square estimate*

| Est. | Industry Code: 264 | | | Industry Code: 221 | | |
|---|---|---|---|---|---|---|
| | Polynomial | Exponential | Power | Polynomial | Exponential | Power |
| (Intercept) | 1.31 | 1.03 *** | 0.17 | -1.16 *** | 0.77 *** | -0.80 *** |
| | (0.94) | (0.23) | (0.38) | (0.25) | (0.14) | (0.20) |
| $(\log X_1)^1$ | -8.51 | | | 2.48 *** | | |
| | (4.82) | | | (0.46) | | |
| $(\log X_1)^2$ | 17.28 * | | | -0.72 ** | | |
| | (8.23) | | | (0.21) | | |
| $(\log X_1)^3$ | -12.06 * | | | | | |
| | (5.50) | | | | | |
| $(\log X_1)^4$ | 2.78 * | | | | | |
| | (1.25) | | | | | |
| $(X_1)$ | | 0.02 ** | | | 0.05 *** | |
| | | (0.01) | | | (0.01) | |
| $\text{Log}(X_1)$ | | | 0.51 *** | | | 0.90 *** |
| | | | (0.12) | | | (0.07) |
| AIC | 35.46 | 96.54 | | -68.63 | 40.43 | |
| BIC | 44.79 | 101.20 | | -61.68 | 45.65 | |
| Log Likelihood | -11.73 | -45.27 | | 38.32 | -17.22 | |
| Deviance | 4.01 | 27.23 | | 0.40 | 5.58 | |
| Num. obs. | 35 | 35 | 35 | 42 | 42 | 42 |
| $R^2$ | | | 0.34 | | | 0.80 |
| Adj. $R^2$ | | | 0.32 | | | 0.80 |
| SRMSE | **0.4776** | 1.4780 | 1.4835 | **0.1278** | 1.3601 | 1.3657 |
| *** p < 0.001; ** p < 0.01; * p < 0.05 | | | | | | |

***Source:*** *Calculated by the author* from Annual Survey of Industries (ASI) data 2010-11 to 2016-17.

## 5. Results and interpretations:

Most of the studies related to the estimation of production function show a decline in the capital productivity and an increase in labor productivity, mostly because of less capital intensity in Indian industries (Somayajulu & George, 1983). We find no exception for this. Here we have no answer for other Indian states, those who are not major industrial states, where investors increase the capital intensity and follow labor displacing techniques when labor productivity is high. However, we have an answer for these six major industrial states: Tamil Nadu, Karnataka, Gujrat, Maharashtra, Haryana and West Bengal.

Through our result, we found almost all selected industry groups need more or less capital-intensive production. As we already discussed in the beginning that if the value of the additional parameter of capital intensity ($\mu$) is greater than zero, then we can say that the industry needs capital intensive production. Moreover, we can say a higher product per man is obtained by increasing the capital per worker because $\mu>0$. Here, increasingly capital in need of higher labor productivity is likely to induce the manufacturing unit to use more capacity in existence. It reveals that the investors of these selected six states' can increase their capital investment.

From the result, we found the fact that the VES production function is consistent and dominant in some important homogeneous and finer classes of industries under the registered manufacturing sector of India. It lends substantial support to the fact that the VES production function is a more relevant hypothesis for Indian industries compared to the homogeneous production function. The capital intensity is an essential variable which cannot be ignored in explaining productivity (Kazi, 1980a).

Now we compare the additional parameters of capital intensity of VES function and elasticity of substitution between [(capital, labor) as nest and capital intensity] of CES function for all industry groups through a diagram. This diagram helps us to identify that which industry groups need capital intensive production as soon as possible and how much more capital intensity they need to increase for their production. In Figure: 1, it is depicted that X-axis measures the elasticity of substitution from CES and Y-axis measures the additional substitution parameter of capital intensity from VES. Industry codes from the right-end side of the diagram, we ignored because as per the CES estimation, the elasticity of substitution of those industry groups are not theoretically reasonable. Therefore, we only focused on the left side of the diagram. As we see that the industry groups are situated approximately between zero and one as per the X-axis. As per Y-axis, it shows that all industry groups are above the zero point. More height means more capital-intensive production needs.

Through this Figure: 1, we can see that which industry groups need capital intensive production through urgent capital investment.  Industry groups on the right-hand side of the red dotted line have the higher priority than the left-hand side. Those right-hand side industry groups need urgent capital investment as soon as possible for capital intensive production. If those industry groups will not get urgent capital investment, then those industries will fall into further trap of excess capacity. Hence, lower capital creates lower labor productivity, which is likely to induce the manufacturing unit to use lesser capacity in existence.

*Figure: 1.* Industry group-wise comparison between substitution parameters of CES and VES estimates

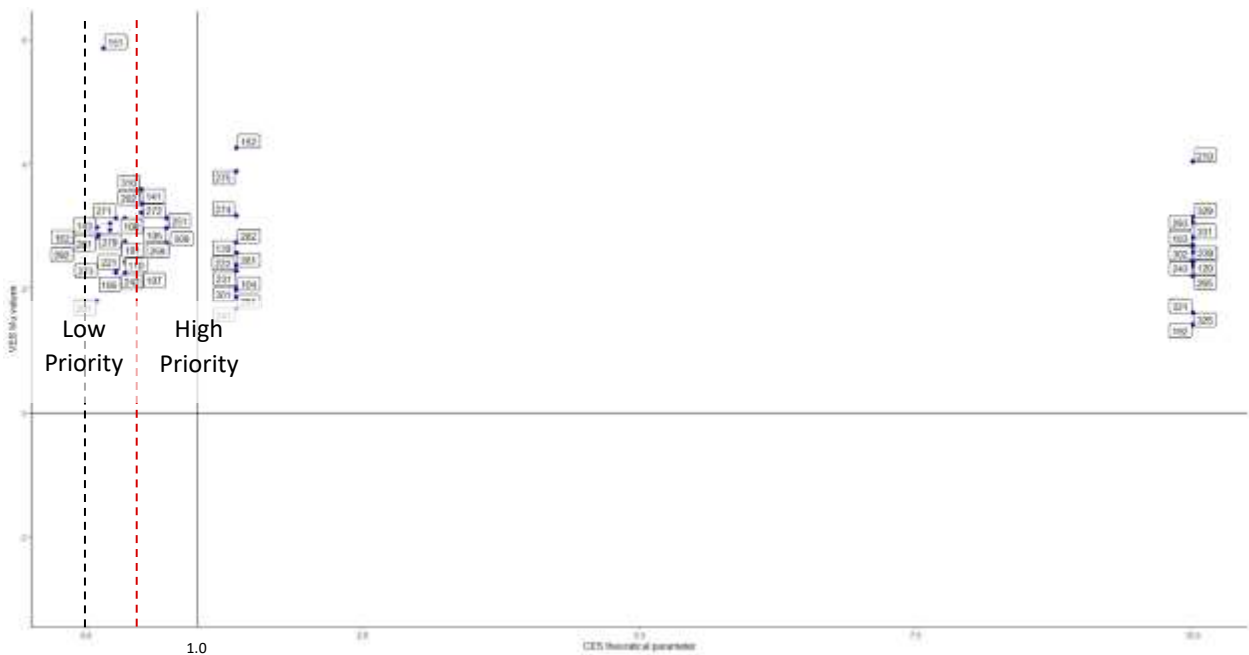

*Source:* Calculated by the author from Annual Survey of Industries (ASI) data 2016-17

## 6.   A comparative exploration of VES function for West Bengal and other states

Industrial units in West Bengal are now facing so many exogenous and endogenous problems that limit their function and force them to run into mediocre performance below the level. Here not all but a few problems are out of the ordinary. Maybe manufacturers would be set with difficulties of distribution, face the rig of the market, get preference changes, storage of power, raw materials, transport, and experience deteriorating industrial relations for many reasons. What may we be able to say, all these are descended from modern manufacturing sector. The business activity where entrepreneurs are to gain to someone in a race to do something, no one will be so callous to suffering. Therefore, their preference reveals to run enterprises with greater excess capacity in which only the addition of technology responds to productivity.

In the Figure: 1, from the high priority zone, we selected three industry groups: 241, 261, and 274 randomly from lower to higher values of µ, respectively. Now we compare the invested capital amount of these three industrial groups state-wise, and then we will describe that which industry groups in which region has the priority to increase their capital investment.

In Figure: 2, we see that for West Bengal industry group 274 has the lowest invested capital amount. Therefore, we can say that the industry group 274 in West Bengal needs urgent capital investment to increase their capacity utilization by capital intensive production. However, this industry group needs more capital investment than the other two groups. For this reason, we decided to take the industry group 274 for exploration. Industry group: 274 is manufacturing electric and lighting equipment. Through the spatial data plots, we see in Map: 1, units or plants of industry group 274 are not optimally distributed in both parts of the West Bengal. In the southern part there are 59 manufacturing plants found, while in the northern part there are no manufacturing plants found for this industry group.

***Figure: 2.*** *Industry group-wise and state-wise total invested capital (2016-17) (Rs. In millions)*

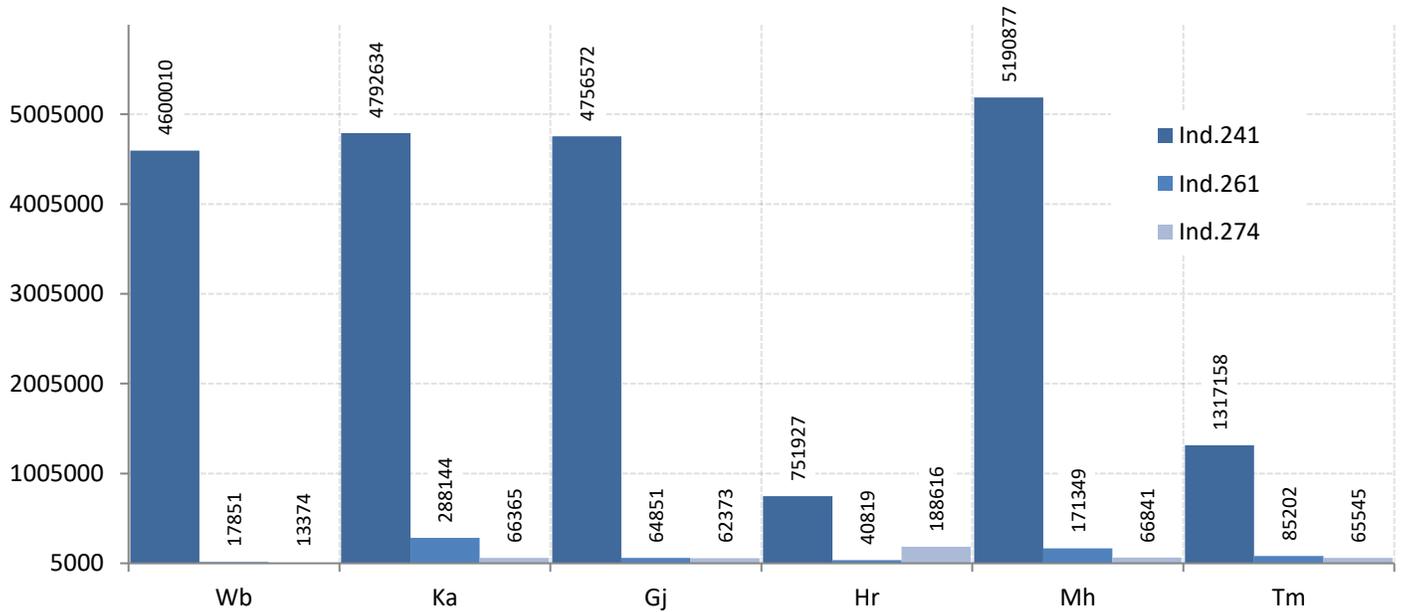

***Source:*** *Calculated by the author from Annual Survey of Industries (ASI) data 2016-17.*

*Map: 1.* *Plots of industrial units or plants of industry group 274 in West Bengal through spatial data analysis*

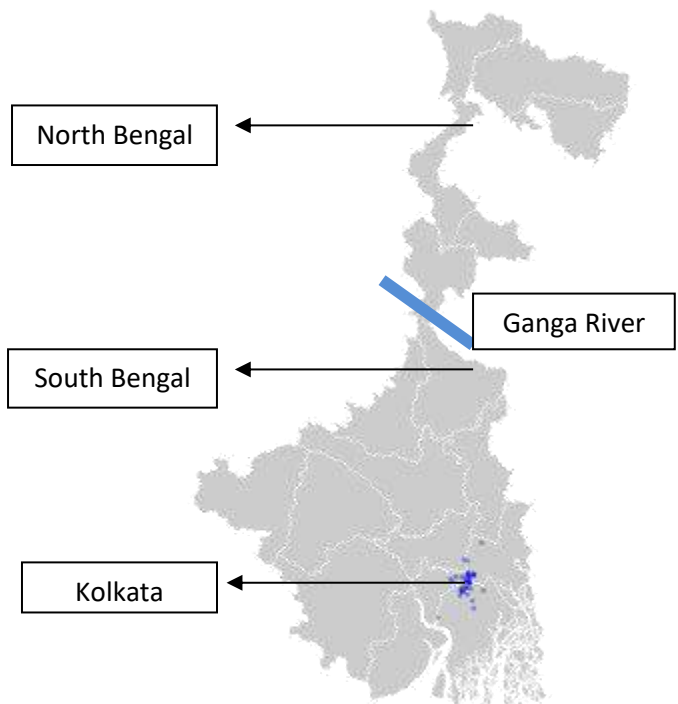

*Source:* *Calculated by the author from Annual Survey of Industries (ASI) data 2016-17.*

The value of μ is 3.17 for this industry group: 274 which is very high. It means that the capital-intensive production method is highly preferred in the state. Therefore, we can partially suggest that if a surplus amount of capital investment from other states of India will be optimally redistributed in West Bengal, then it will increase labor productivity, which will encourage for utilizing the capacity through capital intensive production.

## 7. Policies and Suggestion

We describe that through our results interpretation and exploration study that more capital is needed for gaining the productivity of labor which will induce more capacity of industries in West Bengal. Recently, the government of India started encouraging foreign and domestic investors to invest more in the capital goods of the manufacturing and processing sector through the INVEST INDIA program. To encourage the capital investment, the central government made some orders like Electrical Equipment (Quality Control) Order, 2020 and Public Procurement (Preference to Make in India) Order, 2017. With a mission plan for Indian Electrical Equipment Industry, the central government plans to increase the power generation capacity for the next 5 years. Therefore, technological upgradation is needed for advancing electrical equipment. The central government decided to increase the share of capital goods contribution from 20% to 25% of the total manufacturing activity by 2025. The main aim of the policy is to become one of the top capital goods producing nations in the world (Ministry of Heavy Industries and Public Enterprises, 2016, 2017 & 2020). In National Manufacturing Policy 2011, the central government has announced the objective of enhancing the share of manufacturing in GDP to 25% within a decade and creating 100 million jobs. The policy also emphasizes that it seeks to empower rural youth by imparting the necessary skill sets to make them employable (Ministry of Commerce and Industry, 2011). Therefore, we also suggested that the policy makers of the government of India should focus more on promotion and redesign the labor training scheme (Pradhan Mantri Kaushal Vikas Yojna).

For capital goods investment, the government of India still does not focus attentively. As per our result, we already described that more capital investment is needed for more productivity. On the website of INVEST INDIA program, the central government described that "India's capital goods manufacturing industry serves as a strong base for its engagement across sectors such as engineering, construction, infrastructure and consumer goods, amongst others." If this is true, then as per our CES function, it is identified, industries are still not expanding their technologies. Furthermore, they said that "Capital Goods industry in India provides approximately 1.4 million direct and 7 million indirect jobs." However, as per our literature study, it is concluded that the identified less industrialized states, i.e., West Bengal has still less skilled labor.  Instead of developing and promoting policies, making more new policies with the same content will not cure the problem of excess capacity. For this, our suggestion is that the government of India needs to focus separately for different factors (capital, labor, and capital intensity) in existing policies and promote more efficiently in every corner of the country to optimize the problem of excess capacity.